\documentstyle[12pt,aaspp4]{article}

\newcommand\kms{\ifmmode{~\rm km\th s^{-1}}\else ~km\th s$^{-1}$\fi}
\newcommand\th{\thinspace}

\lefthead{Torres}
\righthead{SS~Lac}

\begin{document}

\title{The change in the inclination angle of the non-eclipsing binary
SS~Lacertae: future eclipses}

\author{Guillermo Torres}
\affil{Harvard-Smithsonian Center for Astrophysics, 60 Garden St.,
Cambridge, MA 02138}
\authoremail{gtorres@cfa.harvard.edu}

\vskip 7in\centerline{To appear in \emph{The Astronomical Journal}, April 2001}

\clearpage

\begin{abstract}

Eclipses in the 14.4 day period double-lined binary SS~Lac were
observed photographically and visually early in the 20$^{\rm th}$
century, but stopped some 50 or 60 years ago. This has been explained
by the presence of a distant third star in the system, which has now
been detected spectroscopically with a period of 679 days. The plane
of the orbit of the binary is changing relative to the line of sight
in response to perturbations from this third object. A recent analysis
by Milone et al.\ of all photometric material available for the
system, including a re-measurement of original Harvard plates, has
confirmed earlier reports of changes in the depth of the eclipses as a
function of time, which are due to the third star.  In this paper we
discuss our detailed analysis of the eclipse amplitude measurements,
and extract from them information on the change in the inclination
angle of the binary over the last century.  Our use of a much improved
ephemeris for the system by Torres \& Stefanik was found to be
crucial, and prompted us to re-determine all the amplitudes from the
historical data at our disposal, including the Harvard material used
by M00.  Systematically lower measurements on the branches of the
minima were properly accounted for, and we made use of both a linear
approximation to the time variation of the inclination angle and a
more realistic model based on the theory of three-body interactions
(``regression of the nodes" effect). The nodal cycle is found to be
$\sim$600~yr, within which \emph{two} eclipse ``seasons" occur, each
lasting about 100~yr.  The non-eclipsing status of the system is
expected to continue until the beginning of the 23$^{\rm rd}$ century. 
	
\end{abstract}

\keywords{binaries: eclipsing --- binaries: spectroscopic --- stars:
fundamental parameters --- stars: individual (SS~Lacertae)}

\clearpage

\section{Introduction}

The cessation of eclipses in a binary star is a rare phenomenon that
can most often be explained by the presence of a third object in the
system inducing perturbations in the orbital elements of the inner
pair. The modulation of the inclination angle to the line of sight, in
particular, is the direct cause of a change in the orientation of the
plane of the binary such that eventually the stars no longer block
each other's light at conjunction.  Few examples of this phenomenon
are known, among them AY~Mus (\markcite{s74}S\"oderhjelm 1974) and
V907~Sco (\markcite{l99}Lacy, Helt \& Vaz 1999). 

An especially interesting case is SS~Lacertae (HIP~108981,
BD$+45\arcdeg$3782, $P = 14.4$~days, \ion{A3}{5}, $\alpha = 22^h 04^m
41\fs6$, $\delta = +46\arcdeg 25\arcmin 38\arcsec$, epoch and equinox
J2000), which stopped eclipsing around the middle of the 20$^{\rm th}$
century.  Our knowledge about light variations in this binary when it
\emph{was} eclipsing comes entirely from visual and photographic
measurements that go back more than a century, but that are
unfortunately of rather poor quality. However, the object has received
considerable attention recently as a result of a re-measurement of
archival plate material and new spectroscopic observations.
\markcite{l91}Lehmann (1991) re-measured original plate material from
the Sonneberg Observatory (1890-1989), presenting the first evidence
that the depth of the eclipses had changed over the years. He
correctly interpreted this as due to the presence of an unseen third
star in the system, which is gradually changing the inclination angle
of the inner pair.  \markcite{tm98}Tomasella \& Munari (1998;
hereafter TM98) derived a double-lined spectroscopic orbit for the
binary, confirming earlier reports that the system had not been
disrupted by a chance encounter with another star and that the period
had not changed, as had been proposed initially to explain the
disappearance of eclipses.  \markcite{m00}Milone et al.\ (2000;
hereafter M00) carried out an exhaustive analysis of all available
light curves of SS~Lac, and presented clear confirmatory evidence of
changes in the depth of the eclipses using their new measurements of
the original Harvard Observatory patrol camera plates (see
\markcite{dw35}Dugan \& Wright 1935; hereafter DW35) as well as other
published measurements.  Further extensive spectroscopic observations
were presented by \markcite{ts00}Torres \& Stefanik (2000; hereafter
TS00), who showed that the system is indeed triple. They determined
the orbital elements of the distant third object, with a period of
about 679~days and a slightly eccentric orbit.  They also detected
apsidal motion for the first time, which is another manifestation of
perturbations caused by the third star.  Further properties and the
history of observations of the system are described in detail by
\markcite{m00}M00 and \markcite{ts00}TS00. 

Adding to its interest, SS Lac is a member of the open cluster
NGC~7209, and therefore other information such as estimates of the
age, distance, and metal abundance is available. Conversely, accurate
dimensions for the binary components, if they can be determined, offer
a chance to perform critical tests of stellar evolution models as well
as valuable clues on the cluster itself, since SS~Lac is located very
close to the turnoff point in the H-R diagram. As pointed out by
\markcite{ts00}TS00, their high-quality spectroscopic orbital solution
for SS Lac has the potential of giving very precise absolute mass
determinations if the inclination angle at the current epoch can be
established accurately enough.  Spectroscopic observations only allow
the minimum mass of each star to be determined, in the form of the
quantity $M \sin^3 i$. The value of the inclination at the present
time can only be inferred by extrapolation, and this requires in
particular that its \emph{rate of change} be known very accurately.
Estimates have been made in a variety of ways by a number of authors,
as described later, but are typically based on very limited data. 

The motivation for this study is precisely to use the new information
that has now become available in the form of light curves analyses and
measurements of the amplitude of the eclipses, to improve our
knowledge of the time dependence of the inclination angle and thus to
allow the absolute masses and also the radii of the component stars to
be established.  We explore ways of doing this and apply a physically
realistic model based on the known properties of the binary and the
effects of the third star expected from theory.  Although as we
describe later this potential can still not be realized because of
limitations in the observations, new insights can be gained on the
timescale and the pattern of the appearance and disappearance of
eclipses. 

\section{Modeling the change in the amplitude of the eclipses}

\markcite{m00}M00 presented measurements of the depth, $D$, of the
eclipses in SS~Lac that show convincingly that the amplitude of the
minima decreased gradually from the 1890's and early 1900's to shortly
before the middle of the 20$^{\rm th}$ century, when eclipses were
last recorded. The peak amplitude is believed to have occurred around
1910 or so. We will re-examine these data in detail in \S3, but we
focus first on a procedure to extract information on the rate of
change of the inclination angle from these measurements. 

As a means of estimating the approximate time of onset and cessation
of eclipses, or the beginning and end of the \emph{eclipse season}, as
they termed it, \markcite{m00}M00 presented a quadratic fit to the
measurements around the primary eclipse as a function of time (their
Fig.~7) that provides a reasonable representation of the data.  They
correctly pointed out, however, that a quadratic fit is not
necessarily a realistic model for the true variation, which is more
likely to taper off at either side of the maximum rather than end
abruptly. As an alternative, \markcite{m00}M00 discussed a fit to a
model involving the complementary error function\footnote{Due to a
misprint the equation appearing in \markcite{m00}M00 is not the actual
model they used, which should read $y = a\cdot{\rm erfc}\left\{[(x -
b)/c]^2\right\}$. The quadratic argument makes the function
symmetrical around the maximum. For the benefit of other readers we
note also that the phases listed in their Table~10 and Table~11, while
computed from the \markcite{tm98}TM98 ephemeris, have an integer
number of cycles added (2713) to render them positive.  Also, the
cycle number for their primary minimum \#40 should read 2421.988.},
which has the desired property of decreasing slowly at both ends
although it still lacks any physical basis and is merely a
mathematical tool. 

The true variation of the depth of the eclipses may be separated into
two parts: a function we call $D(i)$ (one for each minimum), which
describes the change in the amplitude as a function of the inclination
angle $i$, and a second function $i(t)$ that describes the time
variation of $i$. This latter function is what we seek to determine
from the observations. We will assume here that the inclination angle
is the only orbital element that changes significantly with time due
to the third star, although strictly speaking this may not be true for
SS~Lac. On the one hand apsidal motion has been detected
(\markcite{ts00}TS00), and the secular variation in $\omega$ (the
longitude of periastron) will alter the shape of the light curve
slightly over long timescales ($\sim$1000~yr; see
\markcite{ts00}TS00). On the other hand the perturbations from the
third star can also cause a modulation of the eccentricity of the
binary (see, e.g., \markcite{ms79}Mazeh \& Shaham 1979), although this
effect has not yet been detected. Neither of these effects will change
the amplitude of the eclipses appreciably in this case, so for our
purposes all elements except $i$ may be assumed to be constant. 

Similar arguments to describe the time variation of $D$ appear to have
been used by \markcite{l91}Lehmann (1991) in the first report
presenting evidence of morphological changes in the light curve of
SS~Lac, although few details are given. He correctly attributed the
changes in the inclination angle to perturbations from a third object
in the system, and estimated the timescale for the variation. 
	
\subsection{Determining $D(i)$}

The change in the depth of the eclipses as a function of the
inclination angle can be derived in a fairly straightforward way once
a fit to a representative set of observations has been made, using any
of the current computer codes to analyze a binary light curve. One
simply holds all elements fixed and changes $i$ to produce a family of
synthetic light curves, and then measures the depth of the eclipses
(separately at both minima) on each of these as a function of $i$. A
table can then be produced that represents $D(i)$, and this may be
interpolated for ease of use.  The critical step here is arriving at a
light curve solution from the real observations that accurately
represents the physical properties of the components.  The very fact
that the light curve is changing with time and the relatively poor
quality of the historical brightness measurements of SS~Lac make this
step more difficult than it would seem. 

Light curve analyses of the photometric observations for SS~Lac have
been described in detail by \markcite{m00}M00. They used the
University of Calgary version of the Wilson-Devinney code
(\markcite{w92}Wilson 1992; \markcite{msk92}Milone, Stagg \& Kurucz
1992; \markcite{sm93}Stagg \& Milone 1993; \markcite{k98}Kallrath et
al.\ 1998), and examined the three most complete data sets available:

1. Photographic measurements by \markcite{dw35}DW35 based on Harvard
College Observatory patrol camera plates, covering the interval
1890-1934. These were published only in graphical form as an average
light curve, already folded using the ephemeris by the authors.  A
sizeable fraction of these plates and subsequent plates taken at
Harvard were re-measured by \markcite{sbc96}Schiller, Bridges, \&
Clifton (1996), although these new measures were not used by
\markcite{m00}M00 for their light curve analyses. 

2. Photographic measurements by \markcite{w36}Wachmann (1936) based on
plates from Bergedorf spanning the interval 1924-1932. Low light
levels occur only during the last two years of this interval. 

3. Visual estimates by \markcite{k61}Kordylewski, Pagaczewski, \&
Szafraniec (1961) between 1927 and 1948. 

In all cases the solutions were iterated with fits to the radial
velocity measurements of \markcite{tm98}TM98, under various
assumptions on the relative temperatures, luminosities, and mass ratio
of the components.  In their attempt to avoid smearing of the light
curve due to real changes in shape as a function of time,
\markcite{m00}M00 analyzed these three data sets separately.  As a
result some of them are rather sparse, which adds to the difficulties
imposed by the poor quality of the observations. In addition, there
are some inconsistencies between separate solutions, and in the end
\markcite{m00}M00 adopted their results for the \markcite{dw35}DW35
data set as being more reliable. 

A light curve solution for SS~Lac was also presented by
\markcite{ts00}TS00, who chose to merge the datasets by
\markcite{dw35}DW35 and \markcite{w36}Wachmann (1936) since they
overlap in time. They used the computer program EBOP
(\markcite{e81}Etzel 1981; \markcite{pe81}Popper \& Etzel 1981),
which, although simpler than the Wilson-Devinney code, is quite
adequate for this well-detached system. 

The \markcite{m00}M00 and \markcite{ts00}TS00 solutions show some
significant differences, no doubt a reflection of the difficulties
mentioned above, but also because the \markcite{m00}M00 analyses were
done without the benefit of the new spectroscopic information obtained
by \markcite{ts00}TS00, which only appeared later.  The light ratio
(secondary/primary)\footnote{The star designations we adopt here
follow the photometric convention of referring to the star eclipsed at
phase 0.0 (Min~I) as the ``primary". For SS~Lac this happens to be the
less massive and smaller star. The spectroscopic convention (e.g.,
\markcite{tm98}TM98, \markcite{ts00}TS00) has the names reversed.}
derived by \markcite{m00}M00 is approximately 2.2, while the
spectroscopic measurements by \markcite{ts00}TS00 indicate a value
quite close to unity ($1.06 \pm 0.03$), with much larger ratios being
clearly ruled out by those observations. \markcite{tm98}TM98 seem to
concur with \markcite{ts00}TS00, at least qualitatively, reporting
that the secondary is ``slightly" more luminous than the primary from
the relative depths of the Doppler cores and H$\alpha$ line wings in
their spectra.  Because of the inherent difficulty of determining the
ratio of the radii, $k$, from light curves such as that of SS~Lac in
which the stars are nearly equal and the eclipses only partial,
\markcite{ts00}TS00 incorporated their spectroscopic light ratio as an
external constraint on their light curve solution. In this way they
derived a ratio of the radii of $1.05 \pm 0.02$ (secondary/primary).
The unconstrained solution by \markcite{m00}M00, on the other hand,
leads to a value of $k$ around 1.5, similar to that obtained
originally by \markcite{dw35}DW35 also without the use of an external
constraint.  Such a large value seems somewhat unusual for
main-sequence stars with very similar masses ($M_{\rm sec}/M_{\rm
prim} = 1.0279 \pm 0.0031$; \markcite{ts00}TS00).  We note also that
the maximum amplitude of the minima from the \markcite{m00}M00
solution, setting $i = 90\arcdeg$, is $\sim$0.45~mag (see Fig.~1),
whereas their empirical fit to the measured amplitudes, mentioned
earlier, suggests peak values considerably higher than this.  Several
of the measurements of the depth of the eclipses presented by the same
authors (their Table~10 and Table~11) also reach larger values (as
large as 0.76 in one case) between 1900 and 1910. Similar indications
are seen in the work by \markcite{l91}Lehmann (1991).  Admittedly some
of these measurements suggesting deeper minima have rather large
errors due to the photographic nature of the material, and two of them
by \markcite{m00}M00 at about the epoch of maximum amplitude actually
have much lower values, as do two independent measurements by Lehmann.
Further light may be shed on this issue by examining the individual
brightness estimates for SS~Lac from the Harvard plates as re-measured
by \markcite{sbc96}Schiller et al.\ (1996) (see also
\markcite{m00}M00).  These measurements were kindly provided to the
author by E.\ F.\ Milone. Fig.~2 shows a sequence of those
measurements from 12 Harvard plates taken on the night of 1902 May 30
(JD $2,\!415,\!900$) during a primary eclipse.  The brightness
estimates have been referred, for convenience, to the mean light level
of the system outside of eclipse.  Comparison with the same curves
shown in Fig.~1 gives the clear impression that the primary minimum
was indeed deeper than 0.45~mag, and that the data are considerably
better represented by the \markcite{ts00}TS00 curve.  The solution by
\markcite{ts00}TS00 for $i = 90\arcdeg$ indicates depths of about
0.7~mag, which is close to what is expected for component stars of
roughly equal brightness. 

\placefigure{fig1}

\placefigure{fig2}

The adoption of a representative light curve solution has a rather
significant impact on the results described below.  Although the
\markcite{ts00}TS00 solution is certainly not exempt from criticism,
based on the arguments presented above we have chosen for the present
analysis to adopt that solution as a more consistent representation of
the physical properties of the stars in SS~Lac, pending improvements
that may become possible as we discuss later in \S6.  Using EBOP as
described earlier we have determined the functions $D_I(i)$ and
D$_{II}(i)$ for the depths of the primary and secondary eclipse,
respectively, which do not differ greatly. They are represented
graphically in Fig.~3. 

\placefigure{fig3}

\subsection{Time variation of the inclination angle, $i(t)$}

The simplest approximation, and the one usually adopted, is that of a
uniform variation of the inclination angle with time. As a convenient
expression we adopt 
 \begin{equation}
i(t) = 90\arcdeg - di/dt~ (t-t_{90})~,
\end{equation}
 where $t_{90}$ is the epoch at which eclipses are central ($i =
90\arcdeg$).  In \S5 we describe a more realistic representation but
we show also that, given the quality of the data at our disposal, the
linear expression is a sufficiently accurate description in the time
interval covered by the present data. For other purposes such as large
extrapolations or determining the boundaries of the eclipse season of
SS~Lac, a linear change in $i$ is inadequate (see \S6). 

Several estimates of the slope $di/dt$ have been made in recent years
based on a variety of assumptions:

1. \markcite{l91}Lehmann (1991) used measurements of the eclipse
amplitudes based on photographic plates from the Sonneberg Observatory
in a procedure similar to what we described above, and obtained $di/dt
= 0.18 \pm 0.02$~deg~yr$^{-1}$. He also estimated the eclipses to have
been central in $1911 \pm 3$. Few of the details are given, however,
and the original data are not available except in graphical form. 

2. \markcite{tm98}TM98 assumed typical radii for the stars ($R =
2.25$~R$_{\sun}$) based on an adopted spectral classification
\ion{A2}{5}.  From the linear separation between the components
derived from their spectroscopic orbital solution they estimated the
inclination angle at the end of the eclipsing season to be
$83\arcdeg$.  The authors assumed that the eclipses were central in
1900-1915, and also that they disappeared at about 1960 based on
\markcite{l91}Lehmann's (1991) work.  From this they arrived at $di/dt
= 0.13 \pm 0.01$~deg~yr$^{-1}$.  Extrapolating to the mean epoch of
their radial velocity measurements they estimated $i(1998) \sim
78\arcdeg$. 

3. \markcite{m00}M00 estimated the inclination angle at the mean epoch
of the velocity measurements by \markcite{tm98}TM98 to be $i(1998) =
76\fdg5$, through a procedure involving iterations between separate
spectroscopic and light-curve solutions using the photometric data by
\markcite{dw35}DW35. Adopting an epoch of 1912 for central eclipses,
they then obtained $di/dt = 0.157$~deg~yr$^{-1}$. 

4. \markcite{ts00}TS00 combined the photographic measurements by
\markcite{dw35}DW35 and \markcite{w36}Wachmann (1936) to obtain a
light curve solution giving $i = 87\fdg6 \pm 0\fdg2$, which they
assigned to a mean epoch of $1912 \pm 10$. Based on this analysis and
on their spectroscopic orbit, they estimated the minimum inclination
angle for eclipses to occur to be $i_{\rm min} = 81\fdg6$, and adopted
1951 as the epoch of cessation of eclipses. From this they derived
$di/dt = 0.15^{+0.05}_{-0.03}$~deg~yr$^{-1}$. 

Although all these estimates of $di/dt$ turn out to be rather similar,
they are not independent and the assumptions made are quite varied and
sometimes inconsistent, mostly due to limitations in the observational
material. In the sections that follow we describe our own efforts
based on observations that have not been used before for this purpose. 

\section{The data}

Measurements of the depth of the eclipses in SS~Lac were compiled by
\markcite{m00}M00 from several sources: their own re-measurement of
many of the original Harvard plates used by \markcite{dw35}DW35;
photographic observations published by \markcite{w36}Wachmann (1936),
\markcite{n38}Nekrasova (1938), and \markcite{m93}Mossakovskaya
(1993); visual determinations by \markcite{k61}Kordylewski et al.\
(1961); and more recent photoelectric measurements obtained by them at
the Rothney Astrophysical Observatory and also by
\markcite{m93}Mossakovskaya (1993), after the eclipses stopped.  A
total of 43 estimates of $D$ were collected for the primary minimum,
and 27 for the secondary minimum, over a period of nearly 100 years.
Some of them are averages determined from several individual
brightness measurements. In addition to these, \markcite{l91}Lehmann
(1991) reported estimates of the amplitude of the eclipses from
Sonneberg Observatory plates, but they were published only in
graphical form. Further photographic observations of SS~Lac were made
by \markcite{t65}Tashpulatov (1965) from 1937 to 1955. However, these
data have been called into question by \markcite{m93}Mossakovskaya
(1993), and a re-examination of some of the original plate material by
other investigators has cast similar doubts (see \markcite{m00}M00).
Because of this, and because they contain relatively few measurements
at low light levels, we have not considered the Tashpulatov
observations here.  With regard to the other measurements, for the
purpose of this paper we ignore the minor difference between the
visual and photographic bands, following \markcite{m00}M00. 

The crucial ingredient in the selection of brightness measurements for
estimating the amplitude of the eclipses is the ephemeris used to
establish which observations are near an eclipse.  \markcite{m00}M00
adopted the ephemeris by \markcite{tm98}TM98, Min~I~$ =
2,\!450,\!716.32(\pm 0.15) + 14.41638(\pm 0.00010)\cdot E$, which was
the most recent at the time.  A much more accurate determination is
now available from \markcite{ts00}TS00, based on all published times
of eclipse as well as their new radial-velocity measurements, and
accounting also for light travel time (due to the presence of the
third star) and apsidal motion. The sidereal period ($P = 14.4161471
\pm 0.0000089$~days) is significantly different from the previous
value, and one order of magnitude more precise. Over the time interval
in which the eclipse amplitude measurements of \markcite{m00}M00 are
non-zero ($\sim$1890-1940, or 50~yr), the difference in the periods
alone accumulates to a phase shift of 0.02, almost exactly half of the
width of an eclipse. In extreme cases a measurement regarded by
\markcite{m00}M00 to be near mid-eclipse may actually fall beyond the
limits of eclipse, leading to an underestimate of depth of the
corresponding minimum. This is especially true given that
\markcite{m00}M00 adopted the conservative approach (precisely to
allow for possible errors in the \markcite{tm98}TM98 ephemeris) of
including all brightness measurements they determined to be within
$\pm 0.04$ in phase from the minima, which is actually twice the true
half-width of the eclipses. 

An illustration of this effect can be seen in Fig.~4.  In the top
panel we show 17 visual brightness estimates obtained near primary
eclipse by \markcite{k61}Kordylewski et al.\ (1961) on the night of
1930 November 28 (filled symbols), phased with the ephemeris by
\markcite{tm98}TM98.  This is the longest published sequence of
measurements made on a single night during an eclipse that shows a
clear change in brightness.  The light curve displayed for comparison
purposes is from the solution adopted in \S2.1, with a maximum depth
appropriate for the date of these observations. Also shown with a
different symbol are all measurements by \markcite{w36}Wachmann (1936)
that are near primary eclipse. These observations are confined to a
2-yr interval (mean epoch $\sim$1931.8), as mentioned in \S2.1, that
happens to be very close to the Kordylewski date.  The phase shift is
fairly obvious for both data sets. When the same measurements are
folded with the ephemeris by \markcite{ts00}TS00, on the other hand,
the fit is considerably better (lower panel). We point out,
incidentally, that both the Kordylewski observations and the Wachmann
observations are completely independent of the ephemeris by
\markcite{ts00}TS00 since they were not used by the authors in its
derivation, and thus they serve as an independent check on its
accuracy\footnote{The careful reader may have noticed that the Harvard
observations displayed earlier in Fig.~2 (JD~$2,\!415,\!900$,
$\sim$1902.4) were folded with the \markcite{tm98}TM98 ephemeris, yet
they show good phase agreement with the predicted light curves,
seemingly conflicting with the evidence above. As it turns out, if the
phasing is done instead with the \markcite{ts00}TS00 ephemeris the
result is essentially the same because both prescriptions made use of
the main epoch of primary minimum from \markcite{dw35}DW35 in their
determination, which happens to be the very night of these
measurements.  In the case of \markcite{tm98}TM98 this epoch is
imposed exactly, and for \markcite{ts00}TS00 it enters more weakly as
part of a least squares adjustment. The main difference will then be
simply a phase shift, equal to the residual of that epoch of primary
minimum from the \markcite{ts00}TS00 fit (see their Table~6).}. 

\placefigure{fig4}

Additional phase errors may be incurred because of uncertainties in
the epoch, or even in the geometric elements $e$ (eccentricity) and
$\omega$, which essentially determine the location of the secondary
minimum in relation to the primary through the quantity $e
\cos\omega$. Fig.~1 shows some hint of this effect (lower panel). 

As pointed out by \markcite{m00}M00, it is to be expected that some
measurements will fall on the branches of the minima, and therefore
they will underestimate the true amplitude. To assess the importance
of this effect they plotted the measured eclipse depths as a function
of phase (their Fig.~6). From the fact that not all low amplitude
values showed large displacements from the centers of the eclipses
they concluded that the effect was unimportant.  If the phases are
computed instead with the new ephemeris, the general impression from
such a plot is rather different (see Fig.~5): nearly all low amplitude
values are quite far from mid-eclipse, and several are even beyond the
true limits of the eclipse (c.f.\ Fig.~1). This suggests that some of
the amplitudes may indeed be underestimated, and again that the
contamination from measurements outside of eclipse cannot be ignored. 

\placefigure{fig5}

This prompted us to revise the compilation of eclipse depths by
\markcite{m00}M00 by re-examining all the original data using the new
ephemeris. As a result, our list of amplitudes for the primary and for
the secondary eclipse is somewhat different: some estimates were
removed because they lie outside of eclipse, and new ones were added.
The 591 Harvard measurements used by \markcite{m00}M00 were made
available to us by the senior author, and in addition we considered
the measurements from \markcite{w36}Wachmann (1936),
\markcite{n38}Nekrasova (1938), \markcite{k61}Kordylewski et al.\
(1961), and \markcite{m93}Mossakovskaya (1993). 

For each source we first computed the average magnitude outside of
eclipse. Each measurement during an eclipse was then compared to this
average level to derive the amplitude. In the case of the Harvard data
three observations during primary minimum and one during secondary
minimum were much \emph{brighter} than the average level outside of
eclipse (by up to 0.75~mag), and were rejected.  On nights with
several brightness estimates during eclipse we averaged the data as
well as the dates, except in cases when the sequence of measurements
crossed over from the descending branch to the ascending branch. In
such instances the average brightness will always underestimate the
true amplitude at the average phase, and therefore we split the data
at the center of the eclipse. The uncertainty we assigned to an
amplitude estimate is the standard deviation of a single observation
outside of eclipse, for each source. In cases where several
measurements were averaged, the error was decreased by the square root
of the number of observations.  Finally, we included an estimate in
1991 derived from the Hipparcos measurements (\markcite{e97}ESA 1997),
which was shown by \markcite{ts00}TS00 to be very close to
mid-secondary eclipse. Table~1 and Table~2 list all the measurements
we considered for the primary (40) and the secondary eclipse (45),
respectively. 

\placetable{tab1}
\begin{table}
\dummytable\label{tab1}
\end{table}

\placetable{tab2}
\begin{table}
\dummytable\label{tab2}
\end{table}

Unfortunately the original dates of the measurements by
\markcite{l91}Lehmann (1991) have not been published, and this
prevents us from recomputing the phases to establish whether they can
be incorporated into the analysis. We have therefore chosen not to use
them here.

\section{The algorithm and results}

The procedure we followed consists of fitting the $D(t)$ model
(composed of $D_I(i)$ and $D_{II}(i)$ for the primary and secondary
eclipse, and $i(t)$, as described above) to the eclipse depth
measurements, in order to determine $i(t)$.  Measured amplitudes for
both the primary and secondary eclipses were used simultaneously.
Because many of these estimates are not exactly centered at the
minima, and therefore will always underestimate the true amplitude,
they must first be corrected for this systematic effect.  This was
done as follows. For each date of observation we predicted an
inclination angle using a preliminary estimate of $i(t)$. With this we
computed a synthetic light curve using EBOP, which therefore has an
amplitude appropriate for the date of the measurement. Next we
calculated the phase of the observation using the ephemeris, and we
applied a correction to the measured depth equal to the difference in
magnitude between a point on the curve with the same phase as the
observation, and a point at the center of the eclipse.  Amplitude
measurements displaced in phase by 0.017 or more from the center of a
primary or secondary minimum (half width = 0.02) were not considered,
to guard against possible errors in the ephemeris (period or epoch, or
perhaps even residual variations that have not been accounted for
caused, for example, by the presence of additional stars in the
system) and also because such a small drop in brightness is
essentially within the errors of measurement of the visual and
photographic techniques.  We then carried out the fit for $di/dt$ and
$t_{90}$, and improved our preliminary estimate of $i(t)$. The
procedure is iterative, and at each step the corrections are slightly
different. 

The eclipse ephemeris reported by \markcite{ts00}TS00 was derived by
those authors simultaneously with the apsidal motion and with the
spectroscopic elements of the eclipsing binary and of the tertiary
star (including light-time effects, which can amount to 1/10 of the
eclipse half-width). Because the apsidal motion depends on the assumed
inclination angle, and this angle is time-dependent,
\markcite{ts00}TS00 adopted their own determination of $i(t)$ to
account for this variation.  Therefore, the phases we initially
compute here from this ephemeris also depend in principle on that
determination of $i(t)$\footnote{The \markcite{ts00}TS00 determination
of $i(t)$ relied to some extent on observations by
\markcite{t65}Tashpulatov (1965), through the estimated epoch of
cessation of eclipses. Their ephemeris derivation indirectly
incorporated this time dependence of $i$, although in practice it has
little or no effect on the results.  As mentioned in \S3, the
Tashpulatov observations are now considered to be unreliable. However,
because of the procedure adopted here, any residual effect from those
data on the present analysis disappears completely after the first
iteration.}. In order to be self-consistent, for the present analysis
we have recomputed the ephemeris at each iteration using the same
methods and the same data employed by \markcite{ts00}TS00 (times of
eclipse, and radial velocities), starting from their estimate.
Successive iterations use updated expressions for $i(t)$, and the
change in the phases calculated at each step was found to be very
small.  The apsidal motion and spectroscopic elements also did not
change significantly. 

\placefigure{fig6}

The procedure converged in 7 iterations, using standard non-linear
least-squares techniques at each step to solve for $di/dt$ and
$t_{90}$ (e.g., \markcite{p92}Press et al.\ 1992). In order to ensure
that we converged at the absolute minimum of the $\chi^2$ surface,
rather than at a local minimum, we checked the results using a genetic
algorithm (see \markcite{c95}Charbonneau 1995), which explores all of
parameter space. Fig.~6 displays our best fit to the amplitude
measurements of the primary and secondary eclipse, in which the
resulting time variation of the inclination angle is represented by
$i(t) = 90\arcdeg - 0\fdg1441(81) [t - 1905.6(1.4)]$. The maximum
amplitude (corresponding to central eclipses) according to this fit
was therefore reached somewhere between 1904 and 1907. Table~1 and
Table~2 include the phase displacement of each measurement, the final
amplitude corrections we applied, and the corrected eclipse depths as
used in the least squares adjustments and figures. 

\section{A more realistic description of $i(t)$}

As pointed out earlier, a linear approximation to the time dependence
of the inclination angle may be valid over relatively short time
intervals (i.e., the duration of the observations in our case), but in
reality the change in $i$ is more complex. Perturbations induced by
the third star on the inner pair cause the angular momentum vectors of
both orbits to precess around the total angular momentum vector of the
system, which is fixed in space. The phenomenon is known as the
``regression of the nodes" effect (\markcite{s75}S\"oderhjelm 1975;
\markcite{ms76}Mazeh \& Shaham 1976). One of the observable
consequences is a periodic change in the inclination angle of the
inner orbit that may be expressed conveniently for our case as
 \begin{equation}
\cos i = \cos I \cos\epsilon - \sin I \sin\epsilon \cos\left[\Omega_{90} + {2\pi\over P_{\rm node}}(t - t_{90})\right],
\end{equation}
 in which $I$ is the angle between the invariable plane of the system
(which is perpendicular to the total angular momentum vector) and the
plane of the sky. The angle $\epsilon$ is measured between the orbital
plane of the eclipsing binary and the invariable plane, and the
argument of the cosine term on the far right represents the secular
change in $\Omega$, the longitude of the ascending node of the binary
orbit, with a period $P_{\rm node}$. The angle $I$ is constant, but in
general $\epsilon$ may also vary on long timescales, although this
would be problematic to measure and is likely to be a small effect.
For the purpose of this paper we will consider $\epsilon$ not to vary.
As before, $t_{90}$ represents the epoch at which $i$ reaches
$90\arcdeg$. The angle $\Omega_{90}$ is the position angle of the
ascending node at the same instant. From the condition of central
eclipses $\Omega_{90}$ is given by $\cos \Omega_{90} = \cot I
\cot\epsilon$, and thus the four unknowns in the new expression for
$i(t)$ are $I$, $\epsilon$, $P_{\rm node}$, and $t_{90}$. 

Given the quality of the data available for SS~Lac it is unrealistic
to attempt to solve for all four parameters. However, the nodal period
can be estimated from the theory of three-body interactions.
\markcite{ts00}TS00 gave an order-of-magnitude estimate of 500~yr. An
improved value may be obtained from the expression derived by
\markcite{s75}S\"oderhjelm (1975) (his eq.[27]), which depends not
only on the masses of the three stars (and therefore on the unknown
inclination angles of the inner and outer orbits), but also on the
orbital periods (known), the outer eccentricity (known), and the
relative angle between the orbits as well as the orientation of each
orbit relative to the invariable plane of the system. Several useful
constraints on some of these quantities were discussed by
\markcite{ts00}TS00. For example, from the fact that eclipses no
longer occur the inclination angle of the binary at the present time
must be smaller than $i = 81\fdg6$, based on the size and separation
of the components. In addition, the inclination of the plane of the
outer orbit with respect to the line of sight is expected to be no
smaller than $30\arcdeg$ based on the spectroscopic solution, since
smaller angles would imply that the third star would be massive enough
and therefore bright enough that it would have been detected
spectroscopically, if it is a main-sequence star.  This, in turn,
along with the measured apsidal motion, restricts the range of
possible values for the relative inclination angle between the orbits
($\epsilon_{\rm inner} + \epsilon_{\rm outer}$) to be between about
$24\arcdeg$ and $30\arcdeg$ (\markcite{ts00}TS00).  Though all these
limits are helpful, the various angles involved are still not known
exactly, so we have performed Monte Carlo simulations subject to all
available constraints in order to determine the probability
distribution of $P_{\rm node}$.  The result is shown in Fig.~7.
Periods near 680~yr seem more likely, but the range of possible values
extends from 600 to 700~yr.  \markcite{s82}S\"oderhjelm (1982) has
tested the accuracy of the analytical expression used above for
$P_{\rm node}$ by direct integration of the equations of motion in the
three-body problem.  For stars with orbital and dynamical
characteristics similar to those in SS~Lac he concluded that $P_{\rm
node}$ may be overestimated by that equation by up to 10\% or so. We
may therefore adopt as a representative value for SS~Lac a nodal
period of $P_{\rm node} = 600$~yr. 

\placefigure{fig7}

To arrive at the best fit to the observations we again used a genetic
algorithm to search the parameter space, and found two solutions that
are conjugates of each other (the result of a reflection of some of
the angles around $180\arcdeg$). They represent physically equivalent
scenarios that correspond to direct and retrograde motion of the
binary on the plane of the sky. To determine which of these
configurations holds for this system requires astrometric
measurements, which are not available for SS~Lac.  In both cases the
plane of the orbit of the binary is moving away from $90\arcdeg$ at
the present time, and the consequence for eclipses is the same.
Mathematically the ambiguity is expressed by the inclination angle
numerically increasing or decreasing from $90\arcdeg$ during the
20$^{\rm th}$ century.  Choosing the scenario in which $i$ is
decreasing (as was done implicitly in \S4 and following the usual
convention), we performed least-squares fits as in \S4 to solve for
$I$, $\epsilon$, and $t_{90}$, using the data for both eclipses
simultaneously. The corrections to the measured amplitudes were
adopted from the final iteration in \S4 (see Table~1 and Table~2).  We
obtained $I = 84\arcdeg \pm 10\arcdeg$, $\epsilon = 15\arcdeg \pm
3\arcdeg$, and $t_{90} = 1905.8 \pm 1.5$, which are consistent with
all available constraints.  The angles $I$ and $\epsilon$ are
significantly correlated, and this is reflected in their formal
uncertainties.  Fig.~8 shows this fit (solid lines) along with the
corrected observations. For comparison we show also the solution from
\S4 using a linear model for $i(t)$ (dashed lines), which is seen to
be very similar to the new result and provides an equally good fit.
The epoch of maximum eclipse amplitude is essentially the same in both
solutions. 

\placefigure{fig8}

\section{Discussion}

Even though the fits from both models for $i(t)$ are virtually
indistinguishable given the errors in the observations, the
corresponding predictions for the long-term behavior of the
inclination angle and the implications for the eclipsing seasons of
SS~Lac are very different.  According to the linear model the
inclination angle completes a full revolution in 2500~yr, leading to
eclipse seasons that repeat every 1250~yr. However, a secular change
in $i$ is not what is generally expected from three-body interactions.
Instead, the inclination typically oscillates in a roughly sinusoidal
fashion, the exact shape depending on the angles involved in eq.(2).
The period of this modulation for SS~Lac is about $P_{\rm node} =
600$~yr, as discussed above, which is the length of a full nodal
cycle. The eclipse ``seasons" (defined here as intervals when eclipses
happen) can in general occur once or twice during the cycle. In the
present case the indications are that there are \emph{two} seasons per
nodal cycle, each lasting roughly 100~yr. This is shown in Fig.~9,
where we display the depth of the eclipses (primary and secondary)
over an interval of 1500~yr according to the model fit in \S5, as well
as the behavior of the inclination angle to the line of sight.
Eclipses can take place only for values of $i$ in the shaded area
indicated in the lower panel ($81\fdg6 \leq i \leq 98\fdg4$; see
\markcite{ts00}TS00).  The eclipse seasons recur in pairs, which
combined last for about 350~yr.  The observations during the 20$^{\rm
th}$ century (top two panels) are seen to cover the second season of
one of such pairs, and no further eclipses are expected until shortly
after the year 2200, if the present model is correct.  For comparison
we show also in the bottom panel the trend expected for the
inclination angle according to the linear model (dashed line; \S4),
which is seen to match the slope of the more realistic model very well
at the present time. 

\placefigure{fig9}

The conjugate solution mentioned in \S5 gives an identical pattern of
eclipse seasons, but the function representing $i(t)$ is a reflection
about an axis at $i = 90\arcdeg$ of the curve drawn in the lower panel
of Fig.~9. 

The end of an eclipse season depends in practice on the precision of
the measurements, as argued by \markcite{m00}M00. They concluded on
the basis of their analysis of the amplitudes that, in the absence of
photoelectric data, the eclipses effectively ended between 1937 and
1938. The slower decline resulting from the present fits favors a
somewhat later date around 1945 for the primary eclipses and
$\sim$1950 for the secondary eclipses, if we assume that amplitudes
under 0.1~mag are undetectable by the visual and photographic
techniques applied to SS~Lac. Also, central eclipses occurred some 5
or 6 years earlier than in the \markcite{m00}M00 estimate according to
the fits in \S4 or \S5. 

Though probably closer to the truth than the linear form of $i(t)$,
the more sophisticated ``regression of the nodes" model is not without
its weaknesses. The fit to this model is not as robust as the simpler
approximation, partly because of the lack of amplitude measurements to
the left of the maximum. Instead of two unknowns, three quantities
must be solved for (after fixing $P_{\rm node}$) out of the limited
amount of information contained in the data.  In addition, the errors
given for $I$, $\epsilon$, and $t_{90}$ do not include uncertainties
in the mean light curve adopted from \markcite{ts00}TS00, in the value
of $P_{\rm node}$, or in the constraints used on $\epsilon$, which are
based on the measured apsidal motion as reported by
\markcite{ts00}TS00. All these uncertainties are rather difficult to
quantify. We cannot rule out that, given additional measurements of
the depth of the minima, the detailed shape of the function depicted
in Fig.~9 as well as the pattern of the eclipse seasons described
above could change somewhat, particularly regarding the spacing of the
two seasons in each nodal cycle. 

As it turns out, the inclination angle of the orbital plane of the
binary extrapolated to the mean epoch of the spectroscopic
observations by \markcite{ts00}TS00 ($\sim$1998) remains
ill-determined. From the linear model for $i(t)$ we obtain $76\fdg7$,
while the other model predicts $76\fdg6$, which would seem to indicate
good agreement. Although the formal errors in these quantities are
roughly $1\arcdeg$, the true uncertainty is likely to be larger so
that the absolute dimensions of the components can still not be
computed reliably, especially considering also the possible systematic
errors in the mean light curve (see below). 
	
The main limitation in the present analysis lies in the quantity and
quality of the data, both for the amplitude measurements and the light
curves.  However, improvements can still be made in both areas. It is
possible that useful information on the depths of the eclipses can
still be extracted from the Harvard plates that were not re-measured by
\markcite{sbc96}Schiller et al.\ (1996), as well as from the valuable
plate material obtained at the Sonneberg Observatory and used by
\markcite{l91}Lehmann (1991), but never published. The light curve
solutions might also benefit from these new measurements, and this is
important not only to improve our knowledge of the physical properties
of the components in SS~Lac, but for the determination of the depth
functions $D_I$ and $D_{II}$ as well (\S2.1). 

The most difficult problem faced by all previous investigators
attempting to fit the light curves of SS~Lac has been the fact that
the inclination angle of the binary changes with time. Inevitably this
will affect the solution in subtle ways, and it is likely that several
of the photometric elements are biased to some degree as a result of
the averaging of the measurements during the minima, as the amplitude
changes.  A complication of a similar nature has been noted before by
other investigators concerning the apsidal motion in eccentric
binaries, particularly in systems with short apsidal periods. In this
case, even if an accurate apparent period from a linear ephemeris is
used for the primary, the effect is a distortion and widening of the
secondary minimum because of the changing longitude of periastron. An
example where this distortion is quite important is V477~Cyg. 

The solution to the problem of variable photometric elements would
seem rather obvious, yet the common practice has been to circumvent
the issue by dividing up the data into shorter time intervals, with
the effect that data sets are poorer and the phase coverage more
incomplete, as in SS~Lac.  If the change in the element(s) is linear
to first order, as is often the case, it should not be difficult to
correct for this directly in the computer program used to solve the
light curve, allowing for the variation as a function of time. In the
apsidal motion case, this has in fact been done for the system
mentioned above, V477~Cyg, by \markcite{gq92}Gim\'enez \& Quintana
(1992) using the computer program EBOP. They modified the appropriate
subroutines so that the phase calculated for each observation accounts
for the change in the longitude of periastron using a pre-determined
value of $d\omega/dt$.  Similarly, either EBOP, the Wilson-Devinney
code, or other programs in use could in principle be modified to
account for $di/dt$ in calculating the theoretical flux to be compared
with each photometric measurement, thus referring all observations to
a standard (arbitrary) epoch with a \emph{fixed} inclination angle.
More ambitious investigators might be tempted to add $di/dt$ as a
variable to be determined from the solution along with the other light
elements, but this may only be possible in the best observed systems. 

Alternatively, the individual photographic and visual observations for
SS~Lac may first be corrected along the lines described in \S4 for the
amplitude measurements, and then submitted to the light curve
analysis. This would require no change in the current computer codes.
The present analysis of the eclipse amplitude measurements could then
be repeated once an improved light curve solution becomes available,
and the procedure iterated until convergence. 

\section{Final remarks}

Among the very few eclipsing binaries in which the amplitudes of the
minima have been seen to change, to the point where the eclipses
disappear completely due to the cumulative effect of the perturbations
from a third star in the system, only a handful are as well documented
as SS~Lac. Despite obvious limitations in the observational material
spanning more than a century, this has presented us with a rare
opportunity to learn more about the configuration of the system in the
light of current theoretical ideas about three-body interactions among
stars. In particular, the eclipse amplitude measurements provide a
valuable record of the change in the inclination angle that, with the
possible exception of the succinct report by \markcite{l91}Lehmann
(1991), has not been investigated in detail by applying the direct
constraints offered by these observations. 

The three main contributions of this paper to the study of this
phenomenon are (i) the use of a more accurate ephemeris (from
\markcite{ts00}TS00) to select from the historical record the
brightness measurements that were obtained during an eclipse; (ii) the
application of corrections to the amplitude estimates to account for
the systematically lower values typically measured on the branches of
the minima; and (iii) the consideration of more physically realistic
models than used before for the change in the eclipse depth, based on
the known properties of the system (mean light curve solution) and
expressions for the expected time variation of the inclination angle.
For the latter we have explored both a linear approximation that is
found to be quite satisfactory over the period covered by the
observations, and a more rigorous prescription derived from the theory
of the ``regression of the nodes" effect in triple systems. 

Equally good fits are achieved with both expressions for $i(t)$, but
realistic long-term predictions on the future occurrence of eclipses
in SS~Lac can only be made from the second approach. Our fits suggest
that the eclipse seasons (duration $\sim$100~yr) come in pairs that
repeat after about 600~yr, and that we may expect the current lack of
eclipses to last until approximately the beginning of the 23$^{\rm
rd}$ century. 

\acknowledgments

We are grateful to the referee, Dr.\ E.\ F.\ Milone, for a number of
helpful comments as well as for suggesting the use of, and generously
providing the individual measurements of the Harvard patrol camera
plates made by \markcite{sbc96}Schiller et al.\ (1996).  This research
has made use of the SIMBAD database, operated at CDS, Strasbourg,
France, and of NASA's Astrophysics Data System Abstract Service.

\newpage

\begin{figure}
\plotfiddle{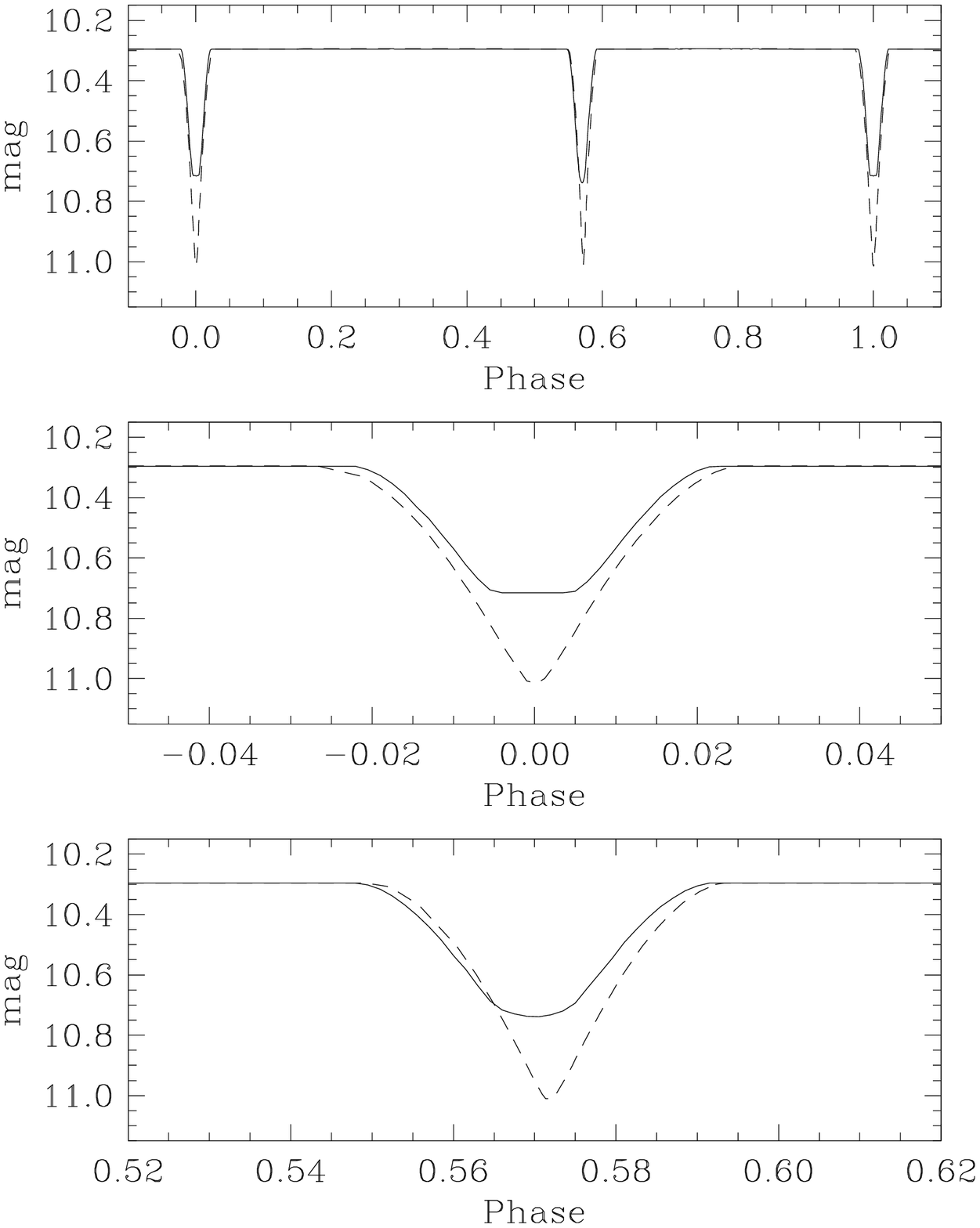}{7in}{0}{85}{85}{-260}{-90}
\caption{Comparison between the light curves
solutions for SS~Lac by M00 (solid line) and TS00 (dashed line), with
the inclination angle set to $90\arcdeg$ to simulate central eclipses
(see text). The lower panels show enlargements near the primary and
secondary minima. The M00 curves were generated from the parameters
listed by those authors and using a standard version of the
Wilson-Devinney code, while the TS00 curve was computed using
EBOP.}
\end{figure}

\clearpage

\begin{figure}
\plotfiddle{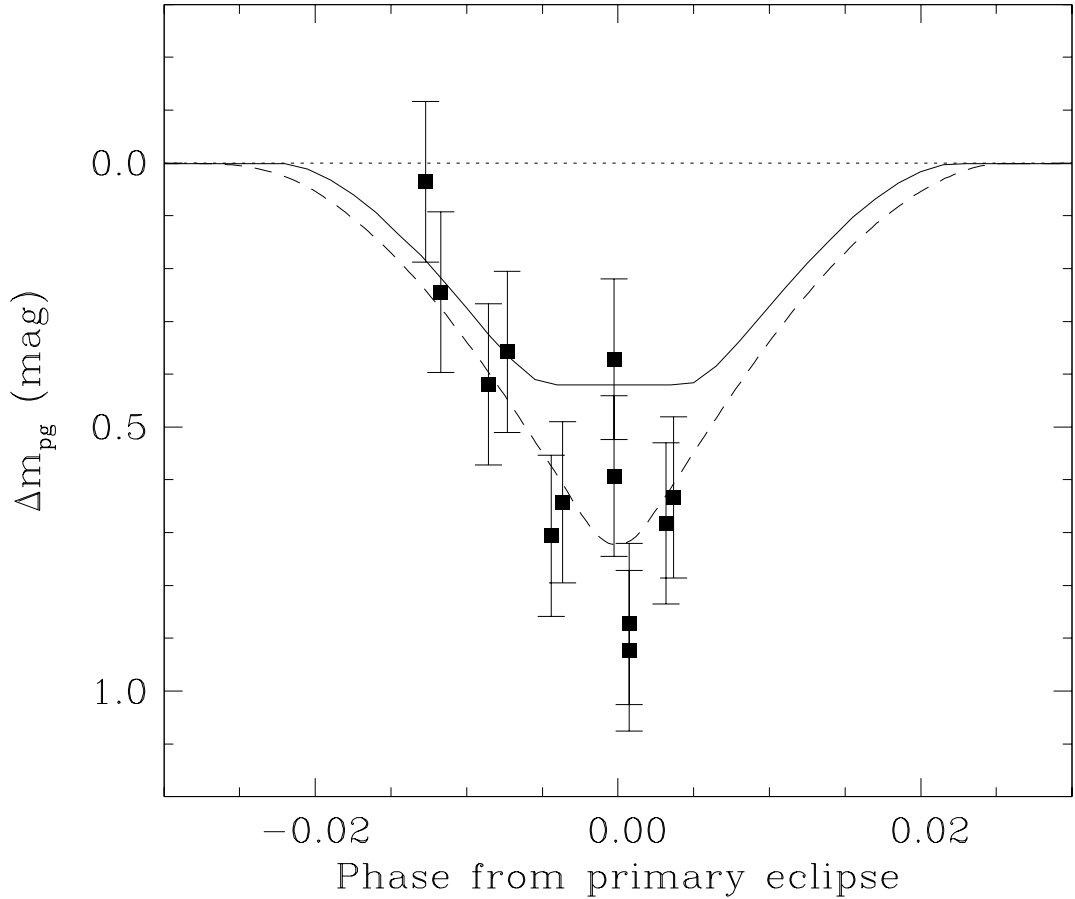}{4in}{0}{85}{85}{-275}{-160}
\caption{Individual brightness measurements for the
night of 1902 May 30 (JD $2,\!415,\!900$) from the Harvard plates (as
re-measured by Schiller et al.\ 1996; see M00), referred to the
average light level outside of eclipse. The error bars adopted
correspond to the uncertainty of a single measurement out of eclipse.
The phasing of the data was done with the ephemeris by TM98. The light
curve solutions displayed for comparison purposes are as in Fig.~1
(M00: solid curve; TS00: dashed curve). The measurements indicate a
primary eclipse amplitude larger than predicted by the M00 curve, and
more consistent with the TS00 solution.}
\end{figure}

\clearpage

\begin{figure}
\plotfiddle{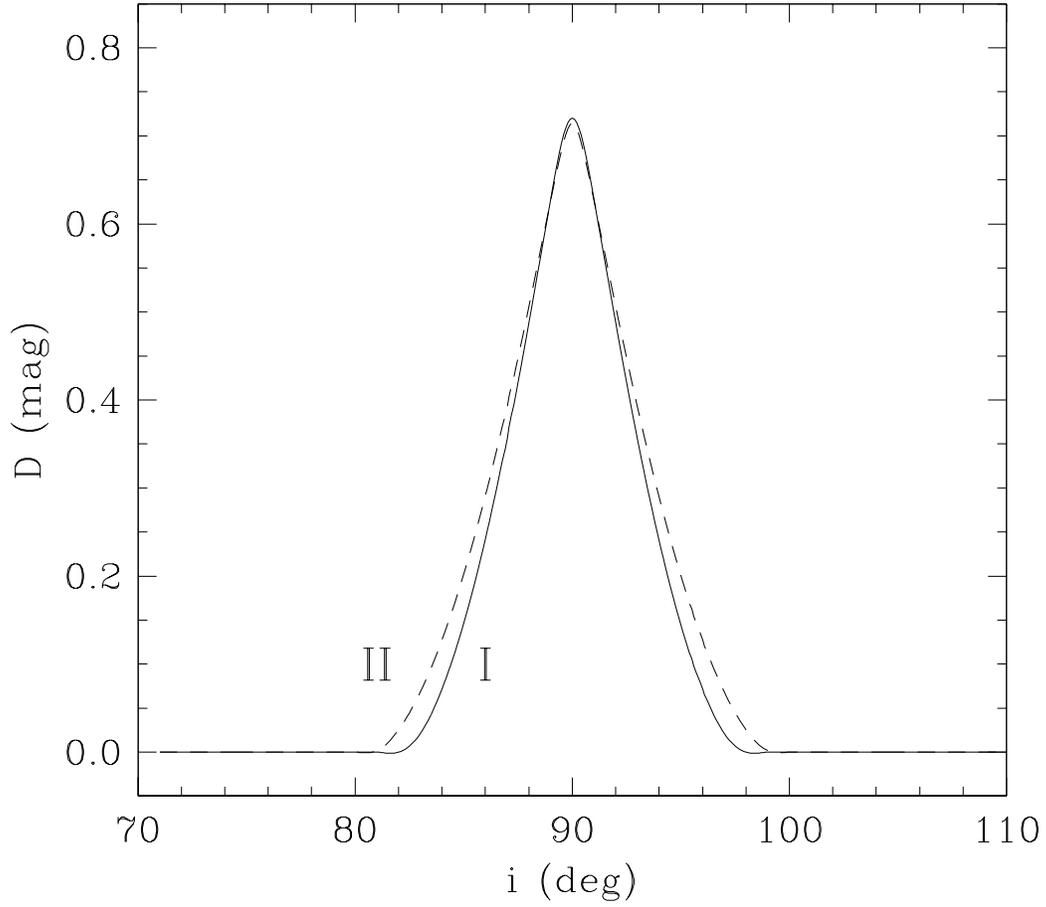}{4in}{0}{85}{85}{-275}{-160}
\caption{Calculated depth of the minima as a
function of the inclination angle, for the primary (I) and secondary
(II) eclipse. The light elements adopted are those of TS00 (see
text).}
\end{figure}

\clearpage

\begin{figure}
\plotfiddle{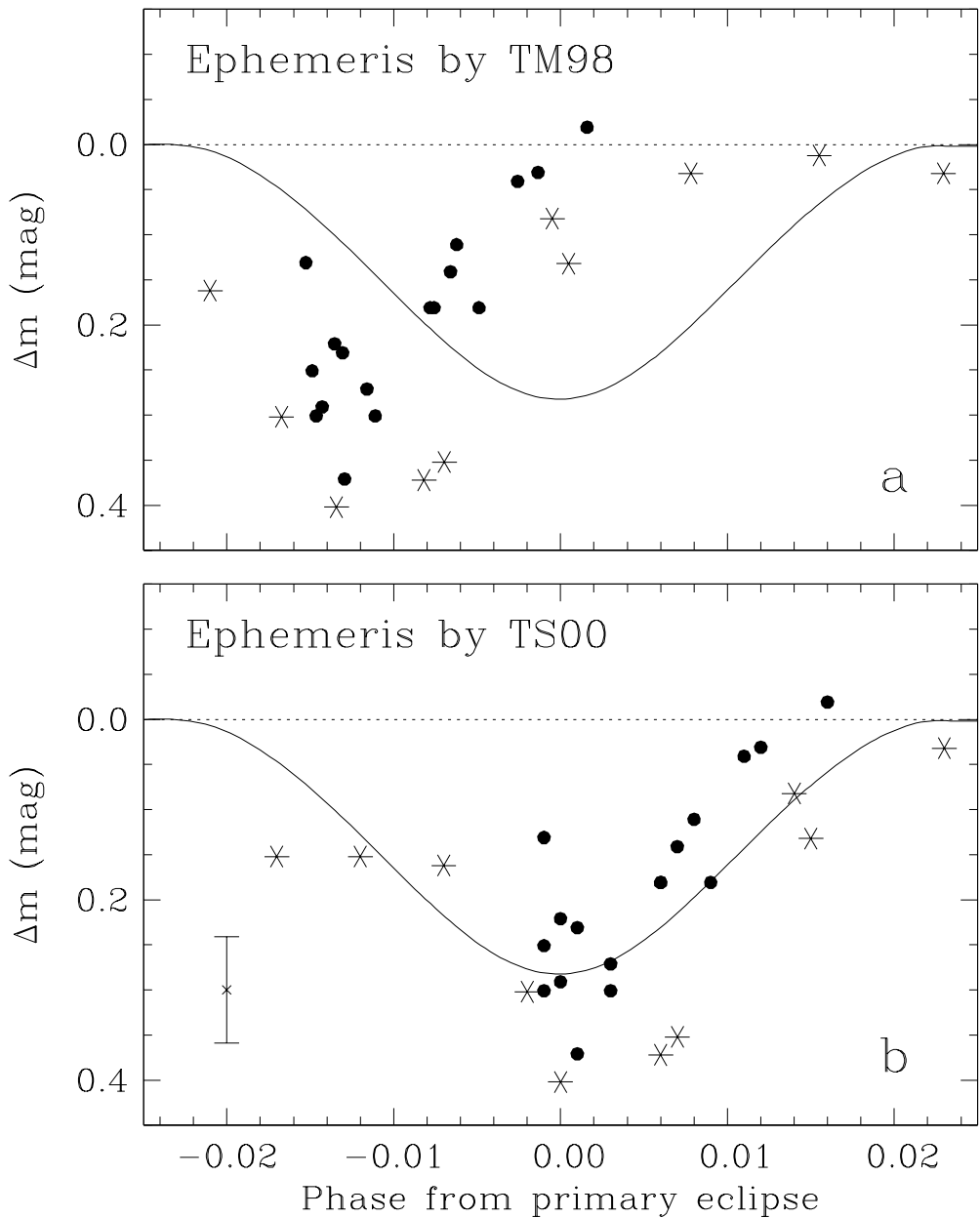}{6.5in}{0}{80}{80}{-260}{-100}
\caption{Visual observations by Kordylewski et al.\
(1961) (filled circles) during primary eclipse on the night of 1930
November 28 (JD $2,\!426,\!309$), and photographic measurements by
Wachmann (1936) from a 2-yr interval with a mean epoch $\sim$1931.8
(stellated symbols). The data are folded (a) with the ephemeris by
TM98, which was adopted by M00, and (b) with the new ephemeris by
TS00.  Magnitude estimates are referred to the mean light level
outside of eclipse. The typical uncertainty is shown in (b).
Displayed for comparison in both panels are synthetic light curves
based on the elements by TS00, with an amplitude appropriate for the
epoch of these observations. The phase discrepancy when using the TM98
ephemeris is apparent.}
\end{figure}

\clearpage

\begin{figure}
\plotfiddle{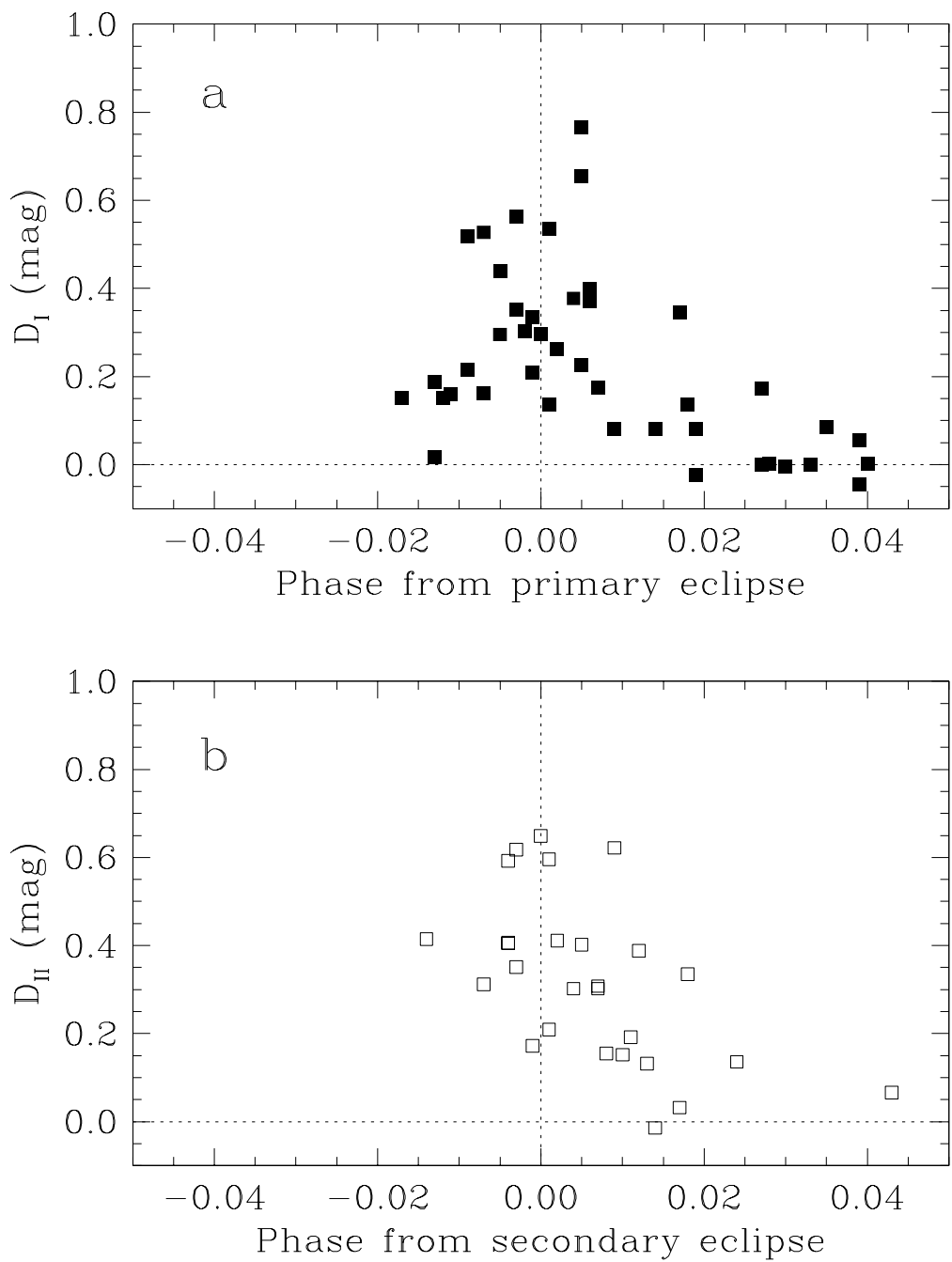}{7in}{0}{85}{85}{-270}{-110}
\caption{Eclipse depth measurements for SS~Lac as
tabulated by M00, shown as a function of phase from the center of the
primary (a) and secondary (b) minimum. The ephemeris used is that of
TS00.  Most low amplitude levels are seen to be significantly
displaced from the center of the eclipses, suggesting that they have
been underestimated because they fall on the branches of the minima.
Some of the points are seen to be beyond the limits of the eclipse,
reflecting inaccuracies in the TM98 ephemeris used by M00 to select
these measurements.}
\end{figure}

\clearpage

\begin{figure}
\plotfiddle{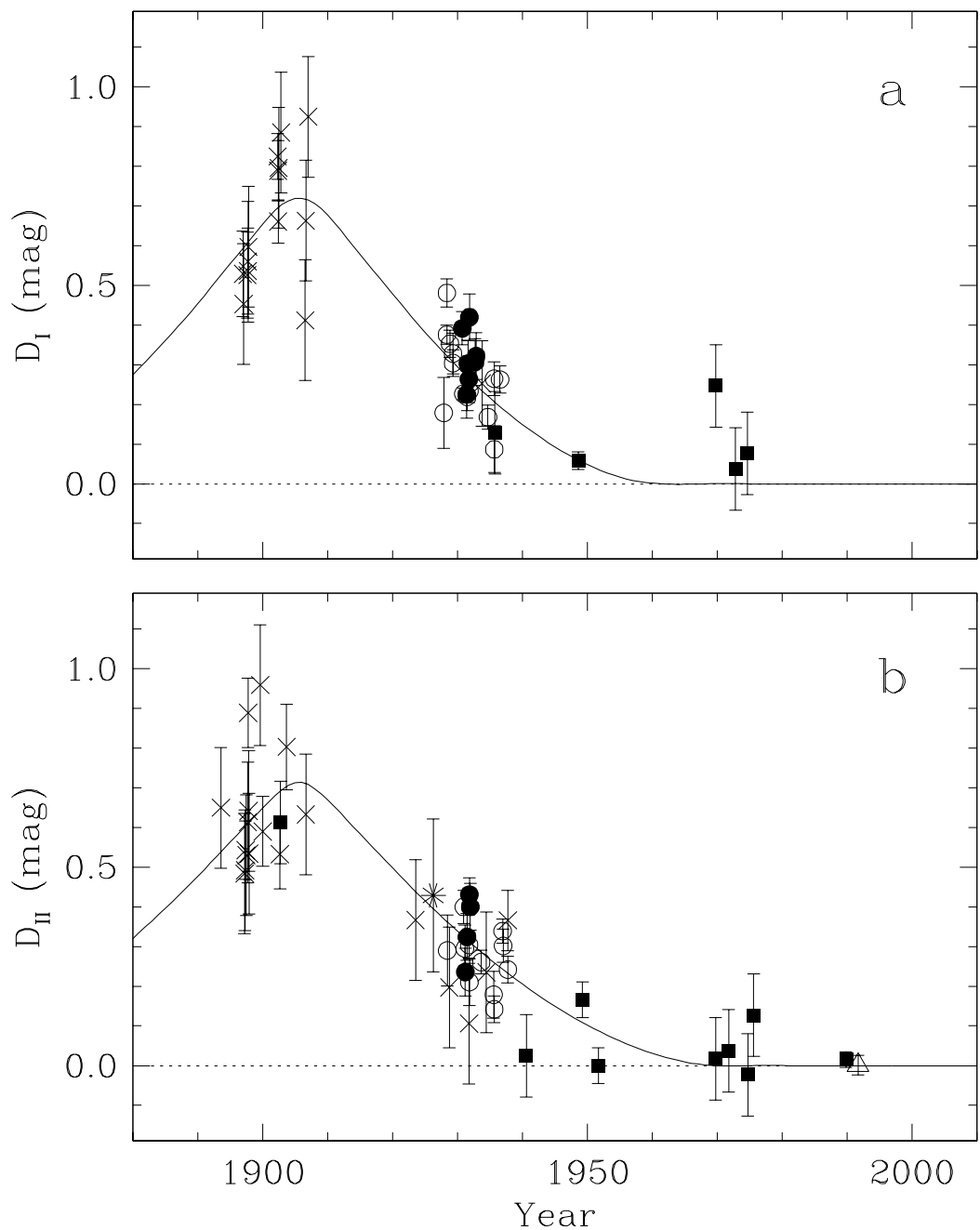}{6.7in}{0}{85}{85}{-270}{-110}
\caption{Fit to the measured eclipse amplitudes
from Table~1 (photometric primary; panel a) and Table~2 (secondary,
panel b), corrected to the center of each eclipse as described in the
text. The fitting model shown here is a combination of the functions
$D(i)$ and $i(t)$, with the latter assumed to be a linear function of
time (eq.(1); see text). The symbols representing different sources
for the measurements are as follows: crosses (Harvard; M00), squares
(Mossakovskaya 1993), asterisk (Nekrasova 1938; a single secondary
measurement), filled circles (Wachmann 1936), open circles
(Kordylewski et al.\ 1961), and open triangle
(Hipparcos).}
\end{figure}

\clearpage

\begin{figure}
\plotfiddle{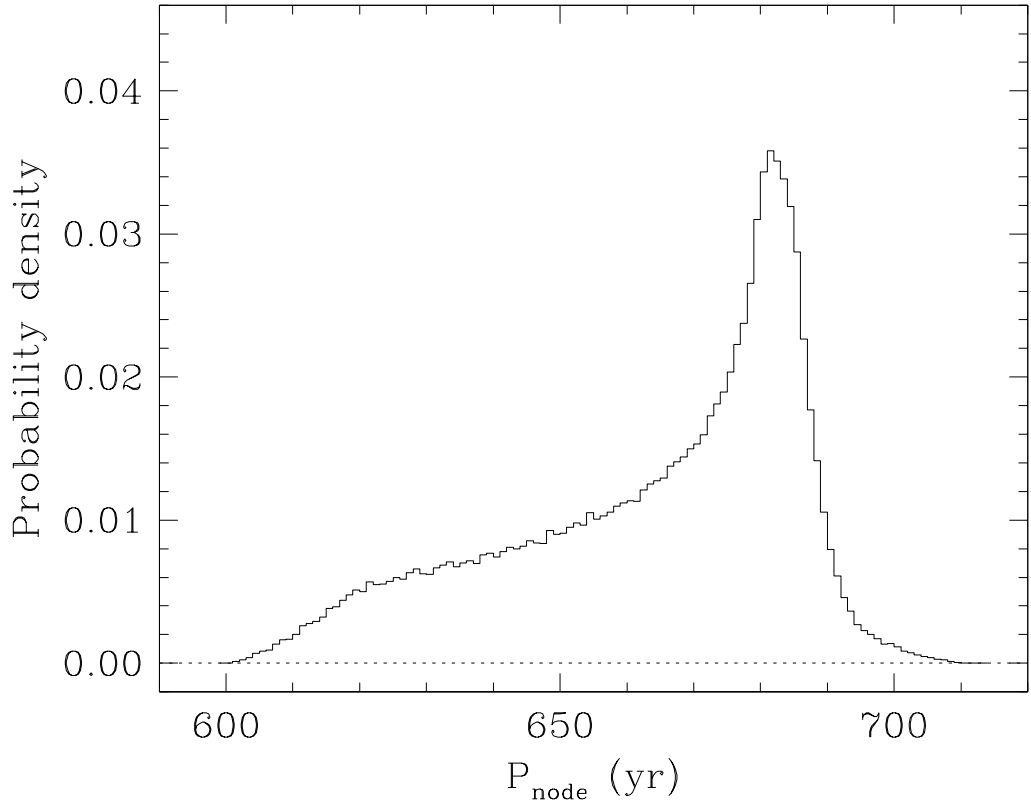}{3.5in}{0}{85}{85}{-270}{-200}
\caption{Probability distribution of the period of
the nodal regression in SS~Lac, from Monte Carlo simulations using
eq.(27) by S\"oderhjelm (1975). All available constraints on the
system have been used, as described in the text.}
\end{figure}

\clearpage

\begin{figure}
\plotfiddle{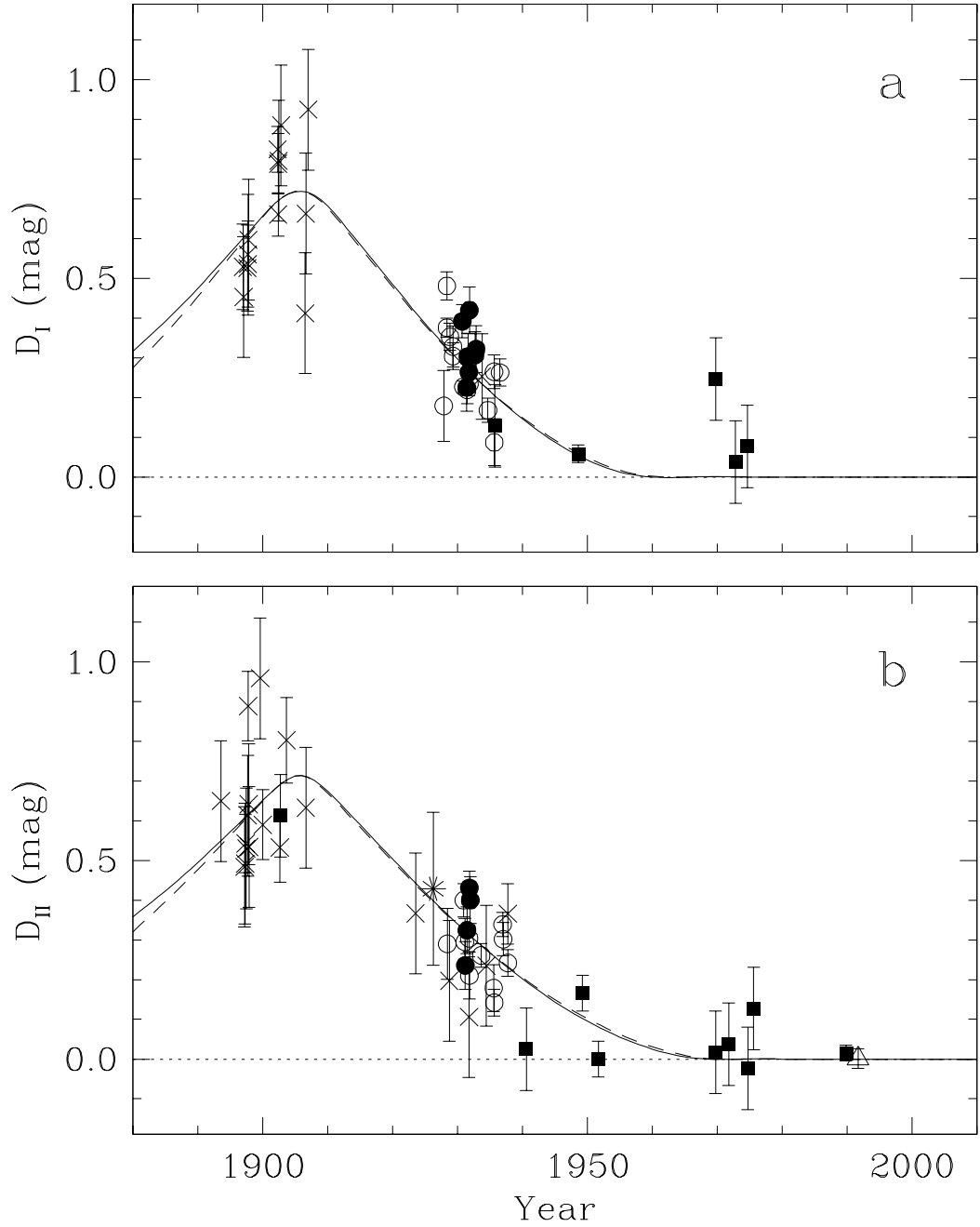}{7in}{0}{85}{85}{-270}{-110}
\caption{Fit to the measured eclipse amplitudes for
the primary (a) and secondary (b), with corrections as described in the
text. The fit represented by the solid lines uses a model for $i(t)$
based on the ``regression of the nodes" effect in triple systems
(eq.(2); see text), rather than a uniform variation, as in Fig.~6. The
latter fit is shown for reference (dashed lines). Symbols for the
individual measurements are as in Fig.~6.}
\end{figure}

\clearpage

\begin{figure}
\plotfiddle{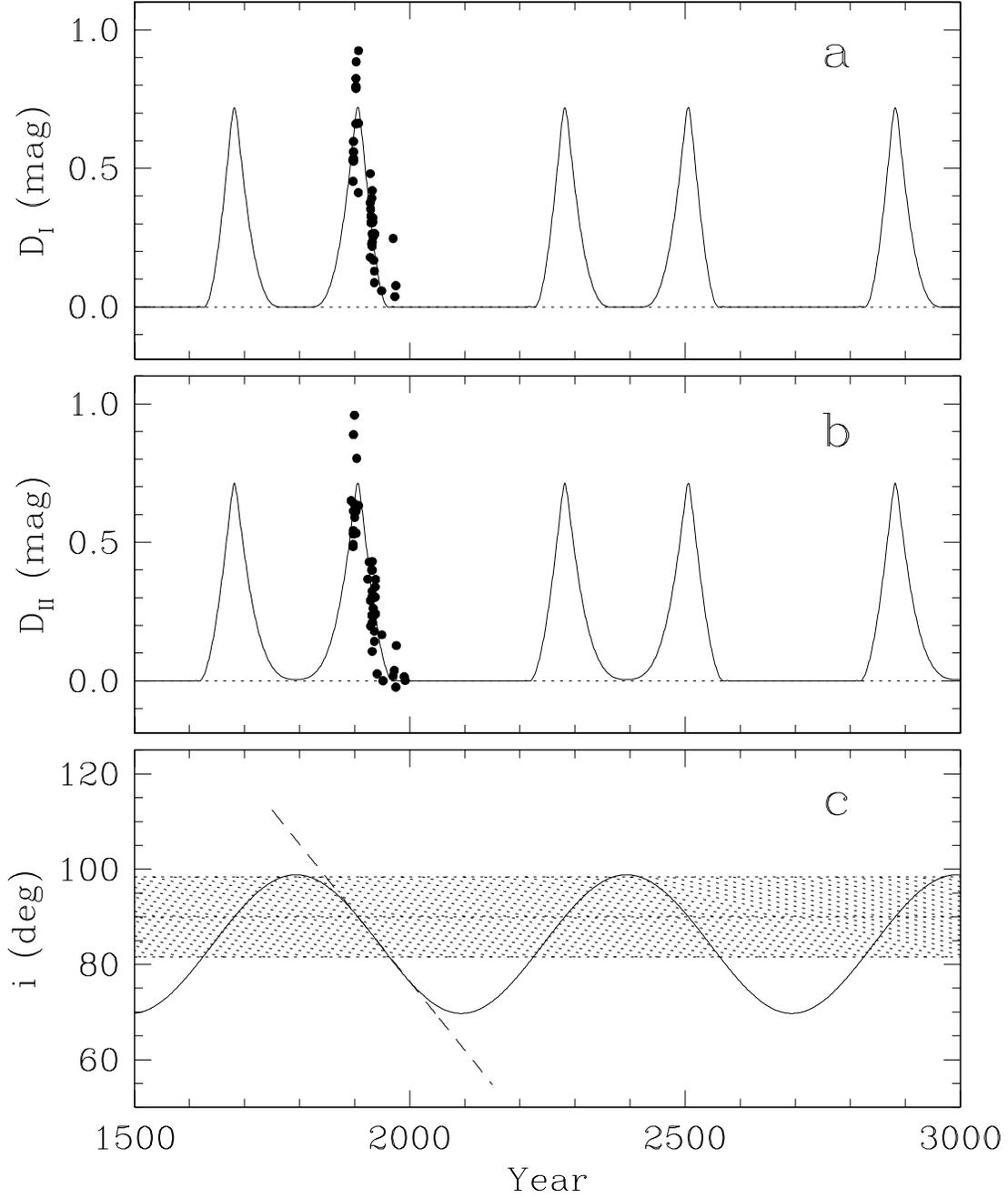}{6.7in}{0}{85}{85}{-270}{-110}
\caption{Long-term behavior of the eclipse
amplitudes for the primary (a) and secondary (b) minima, resulting
from the fit to the measurements based on the ``regression of the
nodes" effect. The cycle repeats with a period $P_{\rm node} = 600$~yr
(see text). Panel (c) displays the expected behavior of the
inclination angle according to this model. The shaded area represents
the range in which eclipses are possible, according to TS00, and the
dashed line shows the behavior based on the linear model for $i(t)$.
Both fits are seen to be essentially indistinguishable over the
interval covered by the observations.}
\end{figure}

\clearpage

\begin{figure}
\plotfiddle{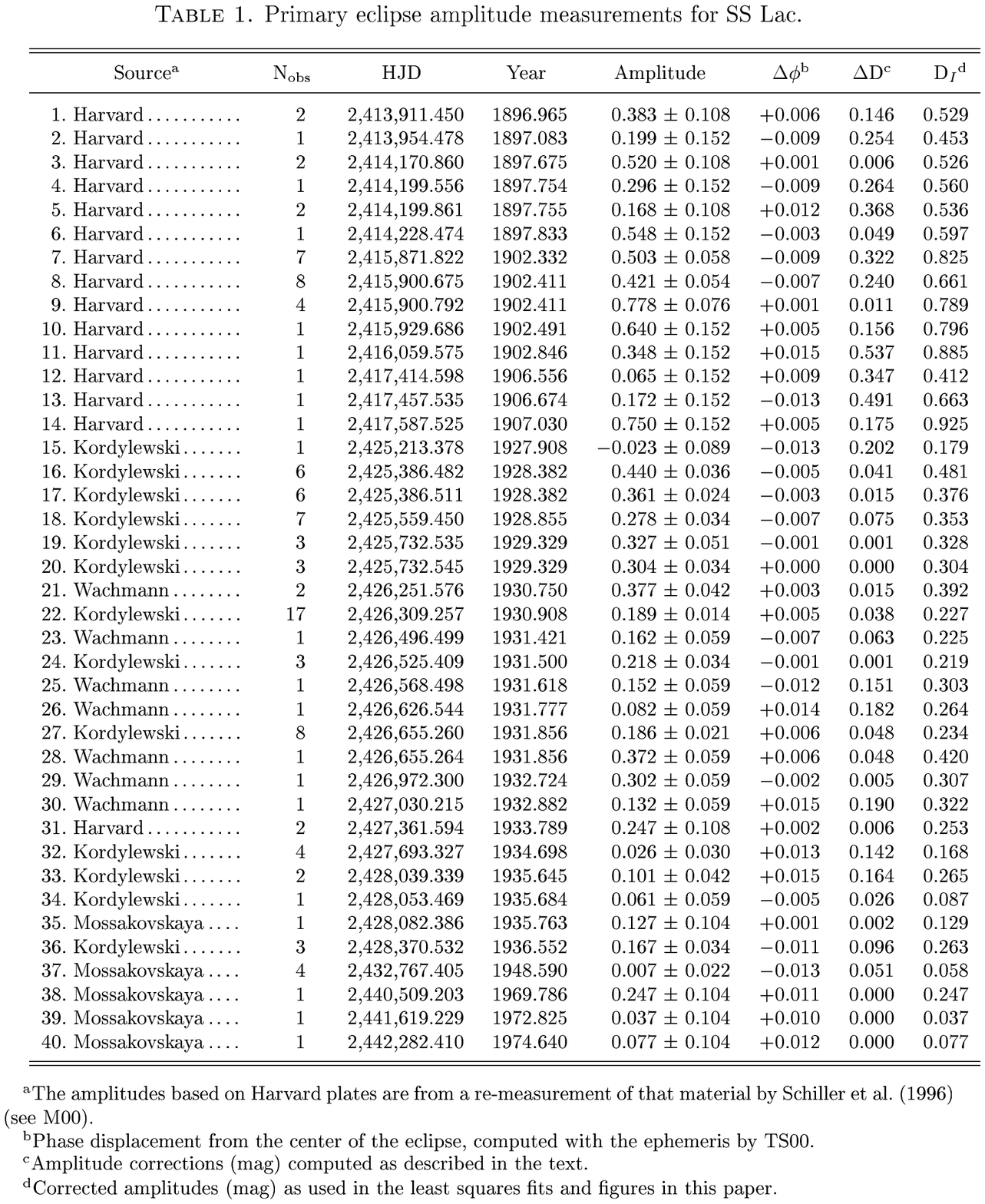}{7in}{0}{90}{90}{-285}{-110}
\end{figure}

\clearpage

\begin{figure}
\plotfiddle{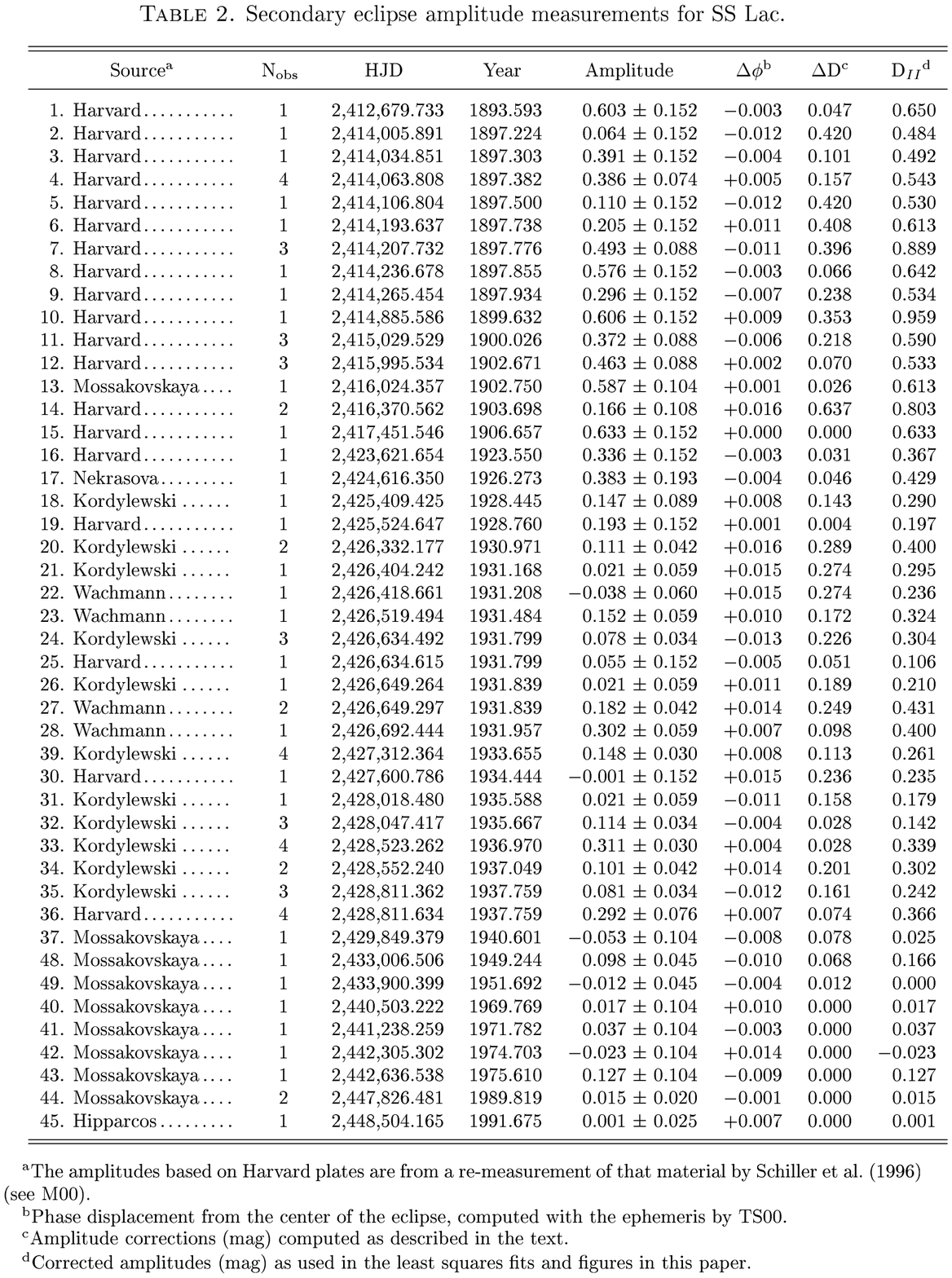}{7in}{0}{90}{90}{-285}{-90}
\end{figure}


\newpage

\begin{references}

\reference{c95} Charbonneau, P. 1995, \apjs, 101, 309

\reference{dw21} Dugan, R.\ S., \& Wright, F.\ W. 1935, \aj, 44, 150
(DW35)

\reference{e97} ESA 1997, The Hipparcos and Tycho Catalogues, ESA
SP-1200

\reference{e81} Etzel, P.\ B. 1981, in Photometric and Spectroscopic
Binary Systems, eds.\ E.\ B.\ Carling and Z.\ Kopal (Dordrecht:
Reidel), 111

\reference{gq92} Gim\'enez, A., \& Quintana, J.\ M. 1992, \aap, 260,
227

\reference{k98} Kallrath, J., Milone, E.\ F., Terrell, D., \& Young,
A.\ T. 1998, \apj, 508, 308

\reference{k61} Kordylewski, K., Pagaczewski, J., \& Szafraniec, R.
1961, Acta Astron.\ Suppl., 4, 487

\reference{l99} Lacy, C.\ H.\ S., Helt, B.\ E., \& Vaz, L.\ P.\ R.
1999, \aj, 117, 541

\reference{l91} Lehmann, T. 1991, Inf.\ Bull.\ Variable Stars, No.\
3610

\reference{ms76} Mazeh, T., \& Shaham, J. 1976, \apj, 205, L147

\reference{ms79} Mazeh, T., \& Shaham, J. 1979, \aap, 77, 145

\reference{m00} Milone, E.\ F., Schiller, S.\ J., Munari, U., \&
Kallrath, J. 2000, \aj, 119, 1405 (M00)

\reference{msk92} Milone, E.\ F., Stagg, C.\ R., \& Kurucz, R.\ L.
1992, \apjs, 79, 123

\reference{m93} Mossakovskaya, L.\ V. 1993, Astron.\ Lett., 19, 35

\reference{n38} Nekrasova, S., 1938, Perem.\ Zvezd., 5, 182

\reference{pe81} Popper, D.\ M., \& Etzel, P.\ B. 1981, \aj, 86, 102 

\reference{pr92} Press, W.\ H., Teukolsky, S.\ A., Vetterling, W.\ T.,
\& Flannery, B.\ P. 1992, Numerical Recipes, 2nd.\ Ed.\ (Cambridge:
Cambridge Univ.\ Press), 650

\reference{sbc96} Schiller, S.\ J., Bridges, D., \& Clifton, T. 1996,
in The Origins, Evolutions, and Destinies of Binary Stars in Clusters,
ASP Conf.\ Ser., 90, eds.\ E.\ F.\ Milone \& J.-C.\ Mermilliod, (San
Francisco: ASP), 141

\reference{s74} S\"oderhjelm, S. 1974, Inf.\ Bull.\ Variable Stars,
No.\ 885

\reference{s75} S\"oderhjelm, S. 1975, \aap, 42, 229

\reference{s82} S\"oderhjelm, S. 1982, \aap, 107, 54

\reference{sm93} Stagg, C.\ R., \& Milone, E.\ F. 1993, Light Curve
Modeling of Eclipsing Binary Stars, ed.\ E.\ F.\ Milone (New York:
Springer), 75

\reference{t65} Tashpulatov, N. 1965, Perem. Zvezd., 15, 424

\reference{tm98} Tomasella, L., \& Munari, U. 1998, \aap, 335, 561
(TM98)

\reference{ts00} Torres, G., \& Stefanik, R.\ P. 2000, \aj, 119, 1914
(TS00)

\reference{w36} Wachmann, A. A. 1936, AN, 258, 361

\reference{w92} Wilson, R.\ E. 1992, Documentation of Eclipsing Binary
Star Model (Gainesville: Univ.\ Florida)

\end{references}
\end{document}